\documentclass[sigconf, nonacm]{acmart}

\newcommand\vldbyear{2026}
\newcommand\vldbworkshop{Applied AI for Database Systems and Applications (AIDB 2026)}
\newcommand\vldbauthors{\authors}
\newcommand\vldbtitle{\shorttitle} 
\newcommand\vldbavailabilityurl{https://github.com/fabian-wenz/rubicon}
\newcommand\vldbpagestyle{plain} 

\usepackage{xcolor}
\usepackage{todonotes}
\usepackage{pifont}
\usepackage[table]{xcolor}
\usepackage{tabularx}
\usepackage{booktabs}
\usepackage{makecell}
\usepackage{dblfloatfix}
\usepackage{diagbox}
\usepackage{soul}
\usepackage{bbm}
\usepackage{amsmath}
\usepackage{booktabs}     
\usepackage{tabularx}     
\usepackage{multirow}     
\usepackage{siunitx}      
\usepackage{colortbl}        
\usepackage{multirow}        

\usepackage{siunitx}

\usepackage{tikz}
\usepackage{pgfplots}
\pgfplotsset{compat=1.18}
\usepgfplotslibrary{groupplots} 
\usepackage{booktabs}
\usepackage{tabularx}
\usepackage{siunitx}
\usepackage{xcolor}
\usepackage{pgfplots}
\usepackage{todonotes}
\usepackage{pgfplotstable}
\pgfplotsset{compat=1.18}
\usepackage{subcaption}

\usepackage{multirow}
\usepackage{natbib}
\usepackage{graphicx}
\usepackage{array}
\usepackage{pifont}
\usepackage{enumitem}

\begin{document}
\title{RUBICON: Agentic AI for Messy Enterprise Data}

\author{
Fabian Wenz$^{1,2}$\quad
Felix Treutwein$^{4}$\quad
Çağatay Demiralp$^{3,2}$\quad
Michael Stonebraker$^{2}$
}
\affiliation{
$^{1}$TU-Darmstadt\quad
$^{2}$MIT\quad
$^{3}$AWS AI Labs\quad
$^{4}$Landeshauptstadt München
}
\email{
fabian.wenz@tu-darmstadt.de,
felix.treutwein@muenchen.de,
{cagatay, stonebraker}@csail.mit.edu
}    
\newcommand{\system}{\texttt{RUBICON}}
\begin{abstract}
Enterprise data exists in many forms, such as tables, text, maps, e-mail, and CAD models, that are invariably access-controlled and behind bespoke interfaces. Current agentic AI systems delegate the entire query workflow to a frontier LLM: a single model is left to interpret the request, decide which sources or tools to use, integrate the evidence retrieved across them, judge its own completeness, and generate an answer--with few constraints, little utilization of schemas, and text as the primary representation throughout. We argue that this methodology is an ineffective abstraction for enterprise data, and that reliable agentic AI should instead require structure: a constrained query interface over each source and a table-centric integration layer that a query processor drives. We introduce \system{}, a system that embodies this vision.

\system{} is based on two fundamental observations. First, text-to-SQL fails on real enterprise data and must be dramatically subsetted to achieve reliable results. Second, data integration across disparate corporate datasets is best performed using tables as the core abstraction rather than the text-centric pipelines of current LLMs and agentic AI systems.

We evaluate \system{} on two benchmarks: on our own enterprise-focused benchmark, \system{}-Bench, against two agentic systems, as well as on a recent semantic query benchmark called SemBench against two state-of-the-art semantic query processing systems, LOTUS and Palimpzest. On \system{}-Bench, where queries require coordinating across heterogeneous enterprise sources, \system{} achieves 100\% end-to-end accuracy, while all agentic baselines — including single- and multi-agent ReAct systems — produce no correct answers. On SemBench, RUBICON surpasses both LOTUS and Palimpzest: it achieves 14.7\% higher accuracy, reduces end-to-end latency by 62.64\%, and lowers token cost by 98.64\%, demonstrating that a table-centric architecture better matches the structure of enterprise data while also yielding significant efficiency gains in enterprise settings.
\end{abstract}

\maketitle

\pagestyle{\vldbpagestyle}
\begingroup\small\noindent\raggedright\textbf{VLDB Workshop Reference Format:}\\
\vldbauthors. \vldbtitle. VLDB \vldbyear\ Workshop: \vldbworkshop.\\ 
\endgroup
\begingroup
\renewcommand\thefootnote{}\footnote{\noindent
This work is licensed under the Creative Commons BY-NC-ND 4.0 International License. Visit \url{https://creativecommons.org/licenses/by-nc-nd/4.0/} to view a copy of this license. For any use beyond those covered by this license, obtain permission by emailing \href{mailto:info@vldb.org}{info@vldb.org}. Copyright is held by the owner/author(s). Publication rights licensed to the VLDB Endowment. \\
\raggedright Proceedings of the VLDB Endowment. 
ISSN 2150-8097. \\
}\addtocounter{footnote}{-1}\endgroup

\ifdefempty{\vldbavailabilityurl}{}{
\vspace{.3cm}
\begingroup\small\noindent\raggedright\textbf{VLDB Workshop Artifact Availability:}\\
The source code, data, and/or other artifacts have been made available at \url{\vldbavailabilityurl}.
\endgroup
}

\section{Introduction}

Large Language Models (LLMs) were all the rage three years ago as researchers tried to figure out what to do with them.  It quickly became apparent that LLMs needed to be augmented by RAG, fine tuning, prompt engineering, etc. to achieve good results.  Hence, LLMs quickly evolved into agentic AI, a pipeline (or graph) of processing steps centered around an LLM.  Agentic AI requires an orchestrator to decide on a sequence of actions to take, and traditional agentic AI entrusts this step to an LLM~\cite{caesura,eckmann2026autonomous}.

We believe there are two fundamental problems with this architecture in an enterprise setting.

First, one needs to access much corporate data through SQL interfaces.  Although LLMs do reasonably well on public text-to-SQL benchmarks, such as \textsc{Spider}~\cite{yu2018spider} and
\textsc{Bird}~\cite{bird} where accuracies above 80\%~\cite{li2024can} are reported, they fail miserably on real-world-inspired benchmarks such as \textsc{Beaver}~\cite{chen2024beaver} (accuracy around 10\%).  Accuracy on Beaver rises to 30+\% when the prompt includes the tables in the \textsc{FROM} clause and all the join terms. The reasons are detailed in \cite{chen2024beaver} and include:
\begin{enumerate}[leftmargin=*, itemsep=0pt, topsep=2pt]
\item \textbf{enterprise data is generally not available for training LLMs as it is proprietary:}  An adage in ~\cite{kandpal} is that an LLM can't retrieve data unless it has seen it at least a couple of times before.

\item \textbf{schema and technical debt ("schema rot"):} This includes opaque names for tables and columns, semantic overlap between tables, and redundancy (materialized views).  This is an example of messy enterprise data. 

\item \textbf{idiosyncratic data:} For example, in the MIT data warehouse buildings have a number, not a name.

\item \textbf{complex queries:}  In real data warehouses the queries are more complex than those in Spider or Bird.
\end{enumerate}
Hence, full SQL is nearly guaranteed to generate failure in an agentic AI environment.  As a result, one requires "baby SQL," for which we propose the Agentic Query Language (AQL).

The second reason concerns the mechanism for data integration.  Using text as the common representation for integration—that is, converting every data source into text and asking an LLM to combine the results—will cost a lot of money in tokens and is likely to produce poor accuracy.  For example, an insurance underwriting application must integrate maps (say of flood zones), CAD (building drawings), previous claims (tables), and underwriting notes (text).  It is very unlikely that converting all these sources to text so an LLM can do the integration will succeed.  A much better idea is to convert the data sources to tables.  This will require a table orchestrator, in other words, a relational query optimizer.

For these two reasons, we propose an alternate agentic AI architecture, well-suited to enterprise data.  It leverages three main concepts:

\paragraph{AQL as the mechanism to access corporate data.} This requires tabular orchestration to generate complex SQL commands, as well as adapters to access data behind bespoke interfaces.
\paragraph{Text access through LLMs, embeddings, or inverted indexes.} The resulting documents are converted to tables and ingested by a query optimizer.
\paragraph{A human in the loop.} Since searches may go ``off the rails,'' the user must be able to inspect intermediate results and correct the plan. Human guidance is also required to orchestrate complex queries.

The remainder of this paper discusses AQL and the architecture of \system{}.  We then conclude with benchmark results comparing \system{} to other approaches on a benchmark inspired by a real application on which we are collaborating with the Department of Transportation of the City of Munich (Germany).

\section{Agentic Query Language}
\label{sec:aql}

\system{} has three moving parts: AQL, wrappers, and intermediate results. AQL is
the user-facing notation. Wrappers are the source-facing translation layer.
Intermediate results are the system-facing representation that makes execution
inspectable and optimizable.

The core abstraction is Agentic Query Language (AQL), a restricted SQL-like
notation that is a compromise between natural language and SQL.  Natural language is
easy to write, but has unacceptably bad accuracy in enterprise settings.  Full SQL has the opposite problem. It is explicit, optimizable,
and reproducible, but requires users to know the schema and write precise
queries. AQL is a compromise between these two extremes.

The basic AQL form is:
\newcommand{\kw}[1]{\textcolor{blue!70!black}{\texttt{#1}}}
\newcommand{\slot}[1]{\textcolor{teal!70!black}{\texttt{#1}}}
\begin{quote}
\texttt{\kw{FIND} \slot{<columns>}}\\
\texttt{\kw{FROM} \slot{<source or table>}}\\
\texttt{\kw{WHERE} \slot{<natural-language predicate>}}
\end{quote}

The user specifies the data source, relation, and desired attributes explicitly; otherwise there is little hope for acceptable accuracy. The
predicate may remain in natural language. AQL also supports different degrees of
explicitness. For example, a novice may write:

\begin{quote}
\texttt{\kw{FIND} \slot{professors}}\\
\texttt{\kw{FROM} \slot{UniversityDW}}\\
\texttt{\kw{WHERE} \slot{faculty members affiliated with the research lab}}
\end{quote}

whereas an expert may write:

\begin{quote}
\texttt{\kw{FIND} \slot{Faculty.name, Faculty.title}}\\
\texttt{\kw{FROM} \slot{UniversityDW.Faculty}}\\
\texttt{\kw{WHERE} \slot{Faculty.affiliation = 'Research Lab'}}\\
\texttt{\slot{AND Faculty.rank IN ('Professor', 'Full Professor')}}
\end{quote}

The attributes specified in the \texttt{FIND} clause are not required to exist in the source. Database
wrappers may construct derived attributes using expressions supported by the
native query language. LLM-derived attributes, however, require an LLM source
whose output is joined with the original relation. For example, in a different query, the LLM may
produce a Boolean attribute \texttt{likely\_about\_AI} for each course, which is then
joined back into the course relation.

Both queries produce the same logical result object:

\begin{quote}
\texttt{ResearchLabProfessors(name, title)}
\end{quote}

The difference is not the desired answer, but how much structure the user chooses
to expose. The more structure is specified, the less work is delegated to the LLM
or wrapper.

AQL also supports joins by composing two source-local queries:

\begin{quote}
\texttt{\kw{FIND} \slot{name, title}}\\
\texttt{\kw{FROM} \slot{UniversityDW.Faculty}}\\
\texttt{\kw{WHERE} \slot{faculty members affiliated with the research lab}}\\
\texttt{\kw{JOIN}}\\
\texttt{\kw{FIND} \slot{name, award}}\\
\texttt{\kw{FROM} \slot{Wikipedia.Pages}}\\
\texttt{\kw{WHERE} \slot{people who have won a Turing Award or Nobel Prize}}
\end{quote}

Here, the shared \texttt{name} attribute supplies the inferred join key. When exact equality is not specified, the join may use entity resolution to match corresponding records across sources.

The important point is not syntactic novelty; rather it is allocation of responsibility. The human or interface specifies the structural part of the query: sources, fields, and intermediate objects. The wrapper or LLM handles only the local ambiguity inside the predicate. This makes LLM use narrow, inspectable, and replaceable.

AQL also includes minimal schema-inspection and housekeeping commands. The
command \texttt{\kw{?}} lists available sources, \texttt{\kw{?} \slot{<source>}} lists available
relations in a source, and \texttt{\kw{?} \slot{<relation>}} lists attributes. Intermediate
results can be named using \texttt{\kw{SAVE}}. These commands are deliberately mundane. Their
purpose is to make exploration explicit rather than conversational.

As a result, a novice can start with mostly natural-language predicates and a GUI that helps select sources and fields, whereas an expert may write more explicit AQL approaching SQL. In both cases, the result is not a hidden chain of tool calls but a visible query object that can be saved, modified, re-executed, and optimized.

An unfamiliar user may initially spend some effort exploring the schema and data. However, a black-box system may require even more investigation if it misinterprets the user’s intent, selects the wrong sources, or returns an incorrect result without revealing where the error occurred. AQL makes this process explicit, allowing the user to inspect and correct a particular source, field, predicate, or intermediate result rather than repeatedly rerunning an opaque end-to-end workflow.

Of course, it is possible to give \system{} the entire user task and ask it to compile the task into an ordered collection of AQL commands. As mentioned earlier, we do not expect \system{} to produce equally reliable results in this mode.

\section{Retrieval as Query Processing}
\label{sec:retrieval-query}

\system{} treats retrieval as a query processing problem. It is closer in spirit to mediator-wrapper data integration than to
RAG-style ingestion: sources remain in place, wrappers expose logical relations,
and integration occurs at query time.

Many enterprise copilots follow an ingestion-and-indexing architecture: they
connect to enterprise systems, extract documents or records, compute embeddings
or sparse indices, and expose a unified semantic search interface. This improves
discoverability, but it does not integrate the underlying data models. A text
index is not a join engine. It is useful for finding evidence; it is not a substitute for an
execution plan.

\system{} instead follows the classical database approach: leave data sources in place, put a wrapper over each source, and perform integration at query time. A wrapper translates AQL into the source’s native interface, relies on the source’s existing execution and access-control mechanisms, and normalizes the result into a logical relation. Hence, \system{} does not reimplement the underlying systems. Adding a source generally requires only a one-time adapter that maps the relevant AQL operations to the source’s API or query language and converts its output into the common tabular representation. The adapter can then be reused across queries. The source may be a database, a search engine, an e-mail system, a web API, or an LLM; the query processor sees each of them through the same logical interface.

For example, an e-mail system can expose a logical table:

\begin{quote}
\texttt{Message(from, to, subject, date, body, thread\_id)}
\end{quote}

A Wikipedia wrapper can expose:

\begin{quote}
\texttt{Page(title, url, snippet, categories, text)}
\end{quote}

A data warehouse already exposes relations, while an LLM wrapper can expose its
answer as a document relation with text, metadata, and provenance. These tables
need not be materialized relations. They are logical views constructed by
wrappers. This normalization step is essential: without a tabular representation, 
downstream operators such as joins, unions, and filters will fail.

Execution proceeds in two modes. In interactive mode, the user issues one AQL
command, inspects the result, and then proceeds. Each command materializes an
intermediate relation that can be inspected, saved, reused, or joined with later
results. This mode is appropriate for exploratory enterprise questions, where the
right plan is often discovered step by step.

In compiled mode, a sequence of AQL commands is executed as a single composite
plan. Once the system has the full command sequence, it can apply database-style
optimization: predicate pushdown, join reordering, source selection, and cost-aware
execution. Costs may include not only runtime, but also tool calls, latency, and
token consumption. In this sense, retrieval is not the step before query
processing. Retrieval is one operator inside query processing.

In the current prototype, source selection and join order follow the user-specified
command sequence. Source-local optimization, including filter pushdown and query
planning, is delegated largely to the wrapped systems before their results are
materialized as intermediate relations. The system records source calls, latency,
and token consumption, but cost-based optimization across heterogeneous sources
remains future work.

In this sense, retrieval is not a step before query processing; it is one operator
within query processing.

\section{RUBICON-Bench and Initial Results}
\label{sec:eval}

We have been working with the Department of Transportation of the City of Munich (Germany). More and more traffic engineers have to spend a significant amount of time answering citizen complaints/queries, often of the form "I don't have time to cross the street before the light turns"  or "I have to wait too long before passing the traffic light at X/Y!" These engineers consult the following data sources to address such concerns:  light schedule (tables), public transport schedule (tables),  diagrams of intersections (CAD), past complaints, and normative context (German laws and Munich specific laws, technical guidelines, etc.).  They have tried agentic AI with helpful qualitative results in simple research tasks on guidelines, but poor results with respect to reproducibility and unsatisfying options regarding the integration of aforementioned sources. They have found the preliminary \system{} results encouraging. Unfortunately, it will be a long process to get permission to anonymize and make public their data and internal benchmarks.

In the meantime, we have constructed \system{}-Bench, which is similar to the Munich problem.  This benchmark models a university-style
enterprise environment in which relevant information is distributed across five
heterogeneous sources: an institutional data warehouse, a research laboratory
website, Wikipedia/API access, e-mail, and LLM knowledge. The data warehouse
contains structured administrative information such as people, buildings, rooms,
and organizational affiliations. The lab website contains semi-structured public
information about research groups, events, projects, and personnel. E-mail
represents private communication, while Wikipedia and the LLM source represent external public knowledge.

The workload contains seven natural-language queries. Each query requires
coordination across two relevant sources, while the remaining sources act as
distractors. Representative queries include:

\begin{quote}
\small

\emph{Which CSAIL professors have won a Turing Award or a Nobel Prize?}


\emph{What research lab events have taken place in a specific campus building
over the past month?}
\end{quote}

These queries are deliberately simple to read but nontrivial to execute. For
example, the question \emph{How many university buildings have a Wikipedia
page?} requires retrieving the set of buildings from the university data
warehouse, retrieving candidate pages from Wikipedia, resolving name variants,
and counting the joined result. Similarly, the question \emph{Which research lab
professors have won a Turing Award or a Nobel Prize?} requires identifying
research lab professors from the institutional source and joining them with
external award evidence. In both cases the
system must use the correct sources and make the cross-source join explicit.


\newcommand{\cmark}{\checkmark}
\newcommand{\dash}{--}
\begin{table}[h!]
\centering
\caption{\system{}-Bench results. Metrics are averaged over queries:
$\overline{T}$ = tokens, $\overline{k}$ = calls, $\overline{C}$ = cost,
Latency = seconds, $Acc1$ = answer accuracy, and $Acc2$ = source accuracy.}
\label{tab:agg_efficiency}
\resizebox{\linewidth}{!}{
\begin{tabular}{l l r rrcrr}
\toprule
\textbf{System} & \textbf{Query} &
{$\overline{T}$} &
{$\overline{k}$} &
{$\overline{C}$ (\$)} &
{Latency(s)} &
{$Acc 1$} &
{$Acc 2$}\\
\midrule
Vanilla & NL        & 1 105.57 & 0.0 & 0.0016 & 13.70 & 0.00\% & 0.00\% \\
Vanilla & AQL       & 1 447.43 & 0.0 & 0.0021 & 12.80 & 0.00\% & 0.00\%\\
\midrule
Single-Agent
& NL        & 7 609.43 & 2.14 & 0.0180 & 16.07 & 0.00\% & 0.00\% \\
Single-Agent
& AQL      & 36 986.00 & 4.57 & 0.0754 & 25.75 & 0.00\% & 71.43\%\\
\midrule
Multi-Agent
& NL        & 34 772.43 & 10.00 & 0.0800 & 47.94 & 0.00\% & 14.29\% \\
Multi-Agent
& AQL         & 8 672.14 & 3.14 & 0.0324 & 28.72 & 28.57\% & 100.00\% \\

\midrule
\textbf{\system} & NL   & 2 440.29 & 2.00 & 0.0048 & 6.99 & 100.00\% & 100.00\%\\
\textbf{\system} & AQL  & 242.29 & 2.00 & 0.0011 & 4.37 & 100.00\% & 100.00\%\\
\bottomrule
\end{tabular}}
\vspace{0.2em}
\begin{minipage}{\linewidth}
\footnotesize
\textit{Note.} All runs in this table use OpenAI's GPT-5.2.
\end{minipage}
\end{table}

Table 1 reports the results of running \system{}-Bench on the execution modes considered in this paper. Each of four systems is run in two different modes. In the natural-language (NL) variant, the system is given only the user question and must decide for itself which sources to access and how to combine them. In the AQL-guided variant, the input is an explicit query plan that names the sources and specifies the intended multi-source execution plan. 

The first system is a vanilla LLM (OpenAI's GPT-5.2), and the two variants are supported through prompt engineering. The next rows show a Single-Agent ReAct~\cite{yao2023react} system which is a single LangChain\footnote{https://github.com/langchain-ai/langchain} ReAct agent with five tools (SQLite, Gmail, Wikipedia, GPT, and CSAIL). Given a natural-language question, the agent freely decides which tools to call and in what order through its own reasoning loop. The AQL-guided variant keeps the exact same agent and ReAct loop but reformats the input as an explicit AQL JOIN plan (naming the two sources and what to retrieve from each).

The third set of rows shows the Multi-Agent variant, namely a LangGraph\footnote{https://github.com/langchain-ai/langgraph} StateGraph where the orchestrator itself is the ReAct-loop\cite{yao2023react}. It reasons ("what do I need next?"), acts (calls one sub-agent), observes the result (inspects collected results), and reasons again before deciding the next step — up to four times. 

Each sub-agent is a single-tool ReAct loop restricted to one data source. The AQL-guided variant uses the same graph but passes the pre-written AQL plan as an instruction, telling the orchestrator exactly which two agents to call and in what order, effectively constraining its iterative reasoning to match the pre-specified plan.

The last option is \system{}. It is evaluated in the same two input modes used throughout the paper. As explained in Section~\ref{sec:aql}, NL mode corresponds to a novice natural-language request, while AQL mode corresponds to a more explicit expert specification. 

We report two accuracy metrics in Table~\ref{tab:agg_efficiency}. $Acc1$ is end-to-end execution accuracy. A query receives score 1 only if the final answer is correct and 0 otherwise. There is no partial credit for calling one of the right sources or making progress toward the answer. On this metric, only \system{} performs well.

$Acc2$ is a source-selection metric. It asks a simpler question: did the system call the correct sources, tools, or agents?  The AQL variants do well on $Acc2$, largely by construction, since the plan specifies the required sources. The NL variants do poorly because the LLM planner must infer the source set from the question alone.

The remaining columns report aggregate efficiency metrics averaged over the benchmark. $\overline{T}$ is the mean number of tokens per query, $\overline{k}$ is the mean number of tool calls, $\overline{C}$ is the mean provider-reported cost, and latency is the mean time to first answer. These metrics show that \system{} has lower latency than the agentic baselines and is generally a great deal cheaper.

The results support several observations. Vanilla LLM scores 0\% on all queries in both modes. Even an explicit AQL sequence does not help. Its single one-shot operation naturally causes low token numbers, cost, and latency. On the other hand, agentic methods result in each additional reasoning step is another LLM call, thereby increasing token consumption, cost, and latency.

Among the agentic baselines, the Single-Agent NL follows a narrow linear chain. It tries to find the answer by successively narrowing the scope, which keeps cost low but usually misses a second data source, if one exists. Of course, Single-Agent AQL fixes the source-selection problem by specifying both sources explicitly. The price is two full ReAct loops plus a synthesis step, which yields roughly five times as many tokens. A Multi-Agent NL system has an orchestrator that sees the full source landscape and explores broadly, exhausting its call budget on every query and producing the highest token count and cost. Multi-Agent AQL restricts execution to the required sources, cutting both tokens and latency.

 $Acc1$ remains near zero for all agent variants. The reason is simple: the sub-agents return prose, not relations. Hence the final ``join'' is a join over unstructured text, which language models do unreliably. \system{}, on the other hand, retrieves relational tables and applies explicit joins. As a result, \system{} is the clear winner across all metrics. Its AQL variant uses the fewest tokens, completes fastest, and achieves 100\% on both $Acc1$ and $Acc2$. The NL variant is, of course, more expensive because of the cost of NL. Even so, it is still cheaper and faster than any agentic baseline, while achieving perfect accuracy.

We also ran \system{} on SemBench~\cite{sembench}, another multimodal query scenario of similar complexity to \system{}-Bench. Notice that \system{}-Bench is focused on whether a system can identify the correct data sources, join their outputs, and determine that the answer is complete.  In contrast, SemBench tests cost and accuracy of fixed pre-specified plans.

We evaluated the SemBench E-Commerce scenario. The workload is one logical source, a fashion dataset, exposed through three access paths: structured attributes, text embeddings, and image embeddings. The joins are therefore intra-catalog joins, not joins across independent enterprise systems. As a result, it is much simpler than Rubicon-Bench.

We executed the workload in the intended \system{} style: source-local \textsc{Find} operations produce intermediate tables, and subsequent operators join these tables explicitly. This simple design works well. \system{} achieves an average quality of 0.86 at a cost of \$0.003 and latency of 55.74\,s. On the same scenario, SemBench reports 0.75 quality, \$0.22 cost, and 149.2\,s latency for LOTUS, and 0.70 quality, \$0.42 cost, and 192.5\,s latency for Palimpzest~\cite{sembench}. For this single-source, multimodal workload, treating semantic predicates as explicit operators and joining materialized intermediate results is cheaper and more robust than pushing the whole task through an LLM-driven execution engine.

\section{Conclusion}
\label{sec:conclusion}

This work has evaluated \system{} against traditional LLMs and agentic AI.  The basic question is whether a less constrained interface can perform well using increased reasoning effort and broader
tool access on multi-source coordination retrieval problems.  Across all
configurations, a consistent pattern emerges: greater autonomy does not 
translate into correct multi-source integration.  Even when equipped with full tool access and high reasoning effort, models
frequently fail to identify all required sources, invoke them in the appropriate
order, or enforce completeness before termination. 

Unless (or until) this problem is rectified, one must use a tool like \system{}.  In addition, one can imagine using a hybrid system where all structured data sources are serviced by something like \system{}.  Then traditional agentic AI would perform the integration between text and tables.  As a next step, we will compare such a system with \system{}, but we are skeptical that it will work better.  Discarding the structure in structured data so it can be combined with text in a text-oriented system just throws away semantics.  It is hard to believe that this is a good idea.  Furthermore, entrusting integration to an agentic AI system increases token consumption and increases cost.  

Overall, the results indicate that LLM-centric coordination does not work well on
enterprise-grade multi-source problems. By contrast, \system{} provides more robust results through its human-in-the-loop, database-oriented architecture.

\bibliographystyle{ACM-Reference-Format}
\bibliography{bibliography}

@article{chen2024beaver,
  author  = {Peter Baile Chen and Fabian Wenz and Yi Zhang and Moe Kayali and Nesime Tatbul and Michael J. Cafarella and {\c{C}}agatay Demiralp and Michael Stonebraker},
  title   = {{BEAVER:} An Enterprise Benchmark for Text-to-SQL},
  journal = {CoRR},
  volume  = {abs/2409.02038},
  year    = {2024},
  doi     = {10.48550/arXiv.2409.02038}
}

@article{bird,
  author  = {Jinyang Li and Binyuan Hui and Ge Qu and Jiaxi Yang and Binhua Li and Bowen Li and Bailin Wang and Bowen Qin and Ruiying Geng and Nan Huo and others},
  title   = {Can LLM Already Serve as a Database Interface? A Big Bench for Large-Scale Database Grounded Text-to-SQLs},
  journal = {Advances in Neural Information Processing Systems},
  volume  = {36},
  year    = {2024}
}

@article{yu2018spider,
  author  = {Tao Yu and Rui Zhang and Kai Yang and Michihiro Yasunaga and Dongxu Wang and Zifan Li and James Ma and Irene Li and Qingning Yao and Shanelle Roman and others},
  title   = {Spider: A Large-Scale Human-Labeled Dataset for Complex and Cross-Domain Semantic Parsing and Text-to-SQL Task},
  journal = {arXiv preprint arXiv:1809.08887},
  year    = {2018}
}

@inproceedings{caesura,
  author    = {Matthias Urban and Carsten Binnig},
  title     = {Demonstrating CAESURA: Language Models as Multi-Modal Query Planners},
  booktitle = {Companion of the 2024 International Conference on Management of Data},
  pages     = {472--475},
  publisher = {ACM},
  year      = {2024},
  doi       = {10.1145/3626246.3654732}
}

@inproceedings{eckmann2026autonomous,
  author    = {Timo Eckmann and Carsten Binnig},
  title     = {A Vision for Autonomous Data Agent Collaboration: From Query-by-Integration to Query-by-Collaboration},
  booktitle = {16th Conference on Innovative Data Systems Research},
  publisher = {www.cidrdb.org},
  year      = {2026},
  url       = {https://vldb.org/cidrdb/2026/a-vision-for-autonomous-data-agent-collaboration-from-query-by-integration-to-query-by-collaboration.html}
}

@article{sembench,
author       = {Jiale Lao and
                  Andreas Zimmerer and
                  Olga Ovcharenko and
                  Tianji Cong and
                  Matthew Russo and
                  Gerardo Vitagliano and
                  Michael Cochez and
                  Fatma {\"{O}}zcan and
                  Gautam Gupta and
                  Thibaud Hottelier and
                  H. V. Jagadish and
                  Kris Kissel and
                  Sebastian Schelter and
                  Andreas Kipf and
                  Immanuel Trummer},
  title        = {SemBench: {A} Benchmark for Semantic Query Processing Engines},
  journal      = {CoRR},
  volume       = {abs/2511.01716},
  year         = {2025},
  doi          = {10.48550/ARXIV.2511.01716}
}

@inproceedings{yao2023react,
  author    = {Shunyu Yao and Jeffrey Zhao and Dian Yu and Nan Du and Izhak Shafran and Karthik R. Narasimhan and Yuan Cao},
  title     = {ReAct: Synergizing Reasoning and Acting in Language Models},
  booktitle = {The Eleventh International Conference on Learning Representations},
  publisher = {OpenReview.net},
  year      = {2023},
  url       = {https://openreview.net/forum?id=WE_vluYUL-X}
}

@article{li2024can,
  title={Can llm already serve as a database interface? a big bench for large-scale database grounded text-to-sqls},
  author={Li, Jinyang and Hui, Binyuan and Qu, Ge and Yang, Jiaxi and Li, Binhua and Li, Bowen and Wang, Bailin and Qin, Bowen and Geng, Ruiying and Huo, Nan and others},
  journal={Advances in Neural Information Processing Systems},
  volume={36},
  year={2024}
}

@inproceedings{kandpal,
author = {Kandpal, Nikhil and Deng, Haikang and Roberts, Adam and Wallace, Eric and Raffel, Colin},
title = {Large language models struggle to learn long-tail knowledge},
year = {2023},
publisher = {JMLR.org},
articleno = {641},
numpages = {12},
location = {Honolulu, Hawaii, USA},
series = {ICML'23},
  doi       = {10.5555/3618408.3619049}
}
\end{document}